\documentclass[english,useAMS]{aa}

\usepackage{graphicx}
\usepackage{amssymb}
\usepackage{psfig}

\def\kms{km\,s$^{-1}$}

\title{VIMOS-IFU survey of $z\sim0.2$ massive galaxy clusters. \\
I. Observations of the strong lensing cluster Abell~2667\thanks{Based 
on observations made with ESO Telescopes at the Paranal Observatories
(programs ID 71.A-3010 and 71.A-0428), 
and on observations made with the NASA/ESA Hubble Space Telescope, 
obtained from the data archive at the Space Telescope Institute. 
STScI is operated
by the association of Universities for Research in Astronomy, Inc.
under the NASA contract NAS 5-26555.
}
}

\titlerunning{VIMOS-IFU observation of ACO~2667}
\authorrunning{G. Covone et al.}

\author{
Giovanni Covone,\inst{1}
Jean-Paul Kneib,\inst{1,2}
Genevi\`eve Soucail,\inst{3}
Johan Richard,\inst{3}
Eric Jullo\inst{1}
\and
Harald Ebeling\inst{4}
}
\institute{
OAMP, Laboratoire d'Astrophysique de Marseille, UMR6110
traverse du Siphon, 13012 Marseille, France
\and 
Caltech-Astronomy, MC105-24, Pasadena, CA 91125, USA
\and 
Observatoire Midi-Pyr\'en\'ees, CNRS-UMR5572, 
14 Avenue E.\ Belin, 31400 Toulouse, France
\and 
Institute for Astronomy, University of Hawaii, 2680 Woodlawn Dr,
Honolulu, HI 96822, USA
}

\date{Received ---, accepted ---.}

\usepackage{babel}
\makeatother
\begin{document}

\abstract{We present extensive multi-color imaging and 
low resolution VIMOS Integral Field Unit (IFU) spectroscopic observations 
of the X-ray luminous cluster Abell~2667 ($z=0.233$).
An extremely bright giant gravitational arc ($z=1.0334 \pm 0.0003$) is easily
identified as part of a triple image system
and other fainter multiple
images are also revealed by the \textit{Hubble Space Telescope} 
Wide Field Planetary
Camera-2 images. The VIMOS-IFU observations cover a field of view of
$54'' \times 54''$ 
and enable us to determine
the redshift of all galaxies down to V$_{\rm 606}=22.5$. 
Furthermore, redshifts could be identified for some sources down to 
V$_{\rm 606}=23.2$.
In particular we identify 21 cluster members in the cluster
inner region, from which we derive a velocity dispersion
of $\sigma=960_{-120}^{+190}$ km s$^{-1}$, corresponding to a total mass
of 7.1$\pm 1.8$\,10$^{13}h_{70}^{-1}$\,M$_\odot$ within a 
$110 \, h_{70}^{-1} $ kpc (30 arcsec)
 radius.  Using the multiple images constraints
and priors on the mass distribution of cluster galaxy halos we
construct a detailed lensing mass model leading to a total mass of 
$2.9 \pm 0.1 \, \times 10^{13}h_{70}^{-1}$\,M$_\odot$ 
within the Einstein radius (16 arcsec). 
The lensing mass and dynamical mass are in good agreement although the 
dynamical one is much less accurate. 
Within a 110~h$_{70}^{-1}$~kpc radius, 
we find  a rest-frame 
K-band M/L ratio of 61$\pm 5 \, h_{70}$\,M$_\odot$/L$_\odot$.
Comparing these measurements with published X-ray analysis, is however
less conclusive. Although the X-ray temperature matches the dynamical and
lensing estimates,
the published NFW mass model derived from the X-ray measurement with its
small concentration of $c \sim 3$ can not account for the large Einstein
radius observed in this cluster. A larger concentration of
$\sim 6$ would however  match the strong lensing measurements.
These results are likely reflecting the complex
structure of the cluster mass distribution, 
underlying the importance of panchromatic studies from small to large scale
in order to better understand cluster physics.
}

\maketitle

\section{Introduction}

\begin{figure*}
\psfig{file=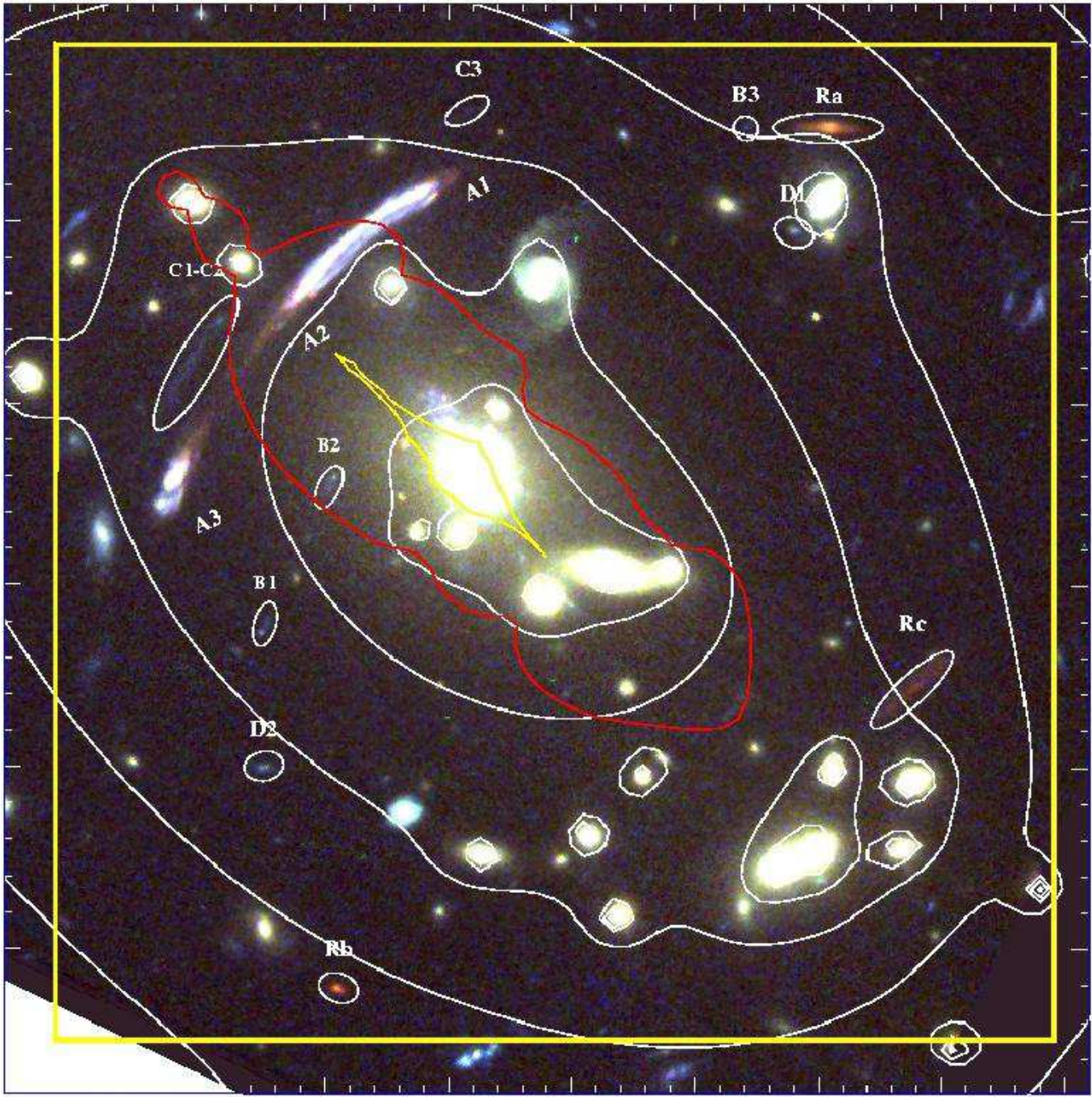,width=1.0\textwidth}
\caption{Color image of the Abell 2667 cluster core imaged with HST in the
F450W, F606W and F814W filters. The thin yellow square represents the position
of the IFU field-of view.
Note the strongly magnified gravitational arc and the
extended blue region just NE of the central galaxy. 
The white lines correspond to
iso-mass contours from the lens model; the red line is the
critical line at the redshift of the giant arc. 
High$-z$ objects discussed in the text are also marked.
North is to the top, East to the
Left. The field of view is centered on $\alpha_{J2000}$=23:52:28.4,
$\delta_{J2000}$=-26:05:08.
At a redshift of $z=0.233$, the angular scale is 3.722 kpc/arcsec. }
\label{fig:color}
\end{figure*}

Massive galaxy clusters are an important laboratory to investigate 
the evolution and formation of the galaxies and the large scale structure of the
Universe 
(see, e.g., the historical review of the study 
of galaxy clusters along the last two centuries in Biviano (2000))
and  
act as natural telescopes magnifying the distant galaxy population. 
Indeed, the most massive clusters are dense enough to have a central
surface mass density larger than the critical density for strong
lensing. This means that multiple images of background sources can
be observed, and in some rare cases, when the source position is close
to the caustics, highly magnified multiple images can be observed
as giant luminous arcs.  It is precisely by the observation of this
rare events that cluster lensing was discovered 
(Lynds \& Petrosian  1986, Soucail et al. 1987, 1988). 
Hence, the discovery of a giant luminous arc in a galaxy cluster is a direct
evidence for a very dense core, and it also corresponds to
a rare opportunity to study in greater details the physical properties
of 
intrinsically faint sources in the high-$z$ Universe
as the magnification factor are generally larger
than 10 (e.g. Ellis et al. 2001, Kneib et al. 2004a, Kneib et al. 2004b, 
Egami et al. 2005).
On the other hand,
using strong and weak lensing in clusters, accurate determination of their mass 
distribution can be compared to numerical predictions of the Cold Dark Matter
scenarios (e.g. Kneib et al. 2003, Sand et al. 2004, Broadhurst et al. 2005).
In some cases, where multiple images with spectroscopic redshift have been
identified in a set of well modeled clusters, it is possible to put
significant constraints on 
the cosmological parameters (Soucail, Kneib \& Golse 2004).

Wide field integral field spectroscopy (IFS)
is a novel observing technique 
with straightforward and important
applications in the observations of galaxy clusters. 
Indeed, IFS provides a tool to obtain in an efficient way
a complete spectroscopic information in a contiguous sky area, 
without the multiplexing difficulties of multi-object spectroscopy, 
and, mostly important, without the need of any {\it a priori} selection
of the targets to be observed.
For instance, this technique can probe rich clusters at $z \sim 1$ 
in the early stages of their formation, to obtain a 
complete (spatially and in magnitude) survey of the galaxy population 
in the inner and more dense regions.

An important application to massive clusters at intermediate redshift
is the survey of the critical lines in the cluster cores
to identify gravitationally lensed objects.
We have thus started an IFS survey of the critical lines in a 
sample of eight massive galaxy clusters using the 
VIsible Multi-Object Spectrograph (VIMOS, Le F\`evre et
al. 2003) Integral Field Unit (IFU), mounted on VLT Melipal, both in 
low and high spectral resolution ($R \sim 200$ and 2500, respectively).
The cluster sample was selected among 
X-ray bright, lensing clusters at redshift $z \simeq 0.2$,
for which \emph{Hubble Space Telescope} (HST) high-resolution imaging 
and {\it Chandra} X-ray observations are already available.
The main scientific goals of this project are the search for 
and the physical characterization of low-mass, highly magnified star
forming galaxies (see, e.g., Campusano et al. 2001), 
the use of strong lensing clusters to constrain
the cosmological parameters,
as explained by Golse, Kneib \& Soucail (2002)
and the study of the 
evolution of the early-type galaxy population in
the rich cluster cores (Covone et al., in preparation).

In this paper, we present the first results from this IFS survey:
novel VIMOS-IFU observations 
and a multi-wavelength imaging analysis of Abell~2667 
($\alpha_{J2000}=23h51m39.4s$, $\delta_{J2000}=-26^{\rm o}05'02''.8$),
a remarkable cluster from the selected sample. 
In this work we focus on the mass model of this cluster using both
strong gravitational lensing and a dynamical analysis
of the cluster core;
in a forthcoming paper we will present a detailed 
multi-wavelength analysis of the 
giant gravitational arc and the other lensed  high$-z$ galaxies.

Abell~2667 is a distance class 6, richness class 3 cluster in the
Abell catalog (Abell 1958), the cD galaxy being 
located at redshift $z=0.233 $.
Although expected to be
an X-ray bright source and to be detected in the ROSAT All-Sky Survey, 
Abell~2667 is not listed in the X-ray flux limited XBACs sample of Ebeling et
al. (1996). The reason for this omission is an unusually large offset
of the X-ray position from the optical cluster centroid listed in
the Abell catalog. 
Abell 2667 was then serendipitously detected 
during  ROSAT PSPC observations of the nova-like star 
V$\star$ VZ Scl in December 1992: the X-ray centroid of the cluster 
was found to be more than 5$'$ off the nominal Abell position.
It is this X-ray observation of Abell~2667 that has motivated observers to include
this object in their optical/X-ray follow-up observations.
In particular, a giant luminous arc was easily identified 
in the cluster core
(Rizza et al. 1998; Ebeling et al., in preparation). 
Remarkably, this arc is so bright that it is detected on DSS2 images 
(although not recognized as an arc).
Using the LRIS spectrograph on the Keck-2 10-m telescope in August
1997 and October 1998, Ebeling et al. (in preparation) 
obtained long-slit spectra of both knots in the giant
arc as well as the third image.
These observations identified the triple arc as a galaxy 
at a redshift of $z=1.034 \pm 0.005$ based on a strong [OII] emission line. 
More recently, a new Keck long-slit optical  
spectrum of this arc has been obtained by Sand et al. (2005),
with the same redshift identification. 

Abell~2667 is among the most luminous galaxy clusters known in the X-ray sky:
ROSAT HRI observations gave an X-ray luminosity\footnote{Throughout this 
paper we will assume a 
cosmological model with $\Omega_{\rm m}=0.3$, $\Omega_{\Lambda}=0.7$
and $H_{0}=70 \, h_{70} \, {\rm km} \, {\rm s}^{-1}$\,Mpc$^{-1}$. 
At $z=0.233$ the angular scale is thus 3.722 kpc arcsec$^{-1}$.}
(in the 0.1-2.4 keV band) of 
$14.90 \pm 0.56 \times10^{44} \, h_{70}^{-2}$ erg s$^{-1}$
and shows a regular X-ray morphology suggestive of a relaxed dynamical state 
(Allen et al. 2003).  Using the same data, Rizza et al. (1998)
estimated a cooling flow time $t_{\rm cool} = 1.8 \times 10^9 $ yrs
(much smaller than the age of the Universe at the cluster redshift), 
and detected the presence of substructure in the intra-cluster medium, 
as evidenced by the shift of $6.0 \pm 2.0$ arcsec
in the X-ray surface brightness centroid position
between the regions in and outside a radius of 150 kpc. 

The brightest cluster galaxy is also a powerful radio source,
with radio flux density $S = 20.1$ mJy at 1.4 Ghz, 
as measured in the NRAO VLA Sky survey (Condon et al. 1998).
This galaxy was also included in the sample 
of radio emitting X-ray sources observed by Caccianiga et al. (2000).
They found strong optical nebular emission lines and classified 
this source as a narrow emission line AGN
(i.e., all the observed emission lines have
FWHM lower than 1000 km s$^{-1}$ in the source rest frame).
More recently, Allen et al. (2003)
have used {\it Chandra} data to
estimate the mass of this cluster: 
by fitting a 
Navarro-Frenk-White (1996; NFW) model,
they have found,
within a virial radius of $r_{200}$,  
%
$M_{200} = 13.6^{+10.6}_{-4.6} \, 10^{14} \, h^{-1}_{70} M_{\odot}$,
with a concentration $c = 3.0 \pm 0.8$.
Fukazawa et al. (2004) have measured an X-ray 
temperature of $6.15 \pm 0.61$ keV (excluding the central cool region),
based on ASCA observations.  
Using archive ROSAT and ASCA, 
Ota \& Mitsuda (2004) have obtained 
a new measurement of the X-ray temperature, found to be
$5.95_{-0.23}^{+0.42} $ keV.

The outline of the paper is the following. 
We present the IFU observations in Sect. 2 along with the
other supporting imaging observations. The IFU data reduction
and analysis is presented in Sect. 3. Sect. 4 presents the 
catalog and the spectroscopic information derived from the IFU data;
Sect. 5 the lensing and dynamical mass models are discussed 
and compared with 
recent mass estimates based on X-ray observations,
Finally, the main results are summarized in Sect. 7.
In the following, magnitudes are given in the AB system.

\section{Observations}

\subsection{Imaging data}

Abell 2667 was observed on October 10-11, 2001 with the 
\emph{HST} using the WFPC2 in the F450W (5x2400 s), F606W (4x1000
s) and F814W (4x1000 s) filters (PI: Allen, proposal ID: 8882). Images
have been retrieved from the ST-ECF archive and reduced using the
\textsc{IRAF} \texttt{drizzle} package (Fruchter \& Hook 2002).
The final spatial
resolution of the images is $0''.05$ per pixel. 
Fig.~\ref{fig:color} shows a
color image of the cluster core made of the 3 \emph{HST} images: 
the giant luminous arc is very prominent, 
and other new candidate multiple images systems 
(see discussion in Sect. 5) are indicated.
The 5$\sigma$ limit detection for point sources on
the final images is 24.76, 25.44 and 24.61 in the filters 
F450W, F606W and F814W, respectively.

On 30 May and 1st June 2003 near-infrared J and H-band observations
with ISAAC have been obtained with the \emph{Very Large
Telescope} (\emph{VLT}) 
(as part of program ID: 71.A-0428, PI: Kneib), 
under photometric sky conditions.
The total exposure time for the J-band and H-band ISAAC data are 
7932 s ($34.8\times4\times57$ s) and 6529 s ($11.4\times10\times57$ s), 
respectively.
The data were reduced using standard {\sc IRAF}
scripts: the final seeing is $0''.51$  and $0''.58$
in the J-band and in the H-band, respectively 
(pixel scale is $0''.148$ per pixel). 
The measured 5$\sigma$
detection limit for point sources are respectively
25.6 and 24.7 in J and H, respectively.

\subsection{VIMOS-IFU 3D spectroscopy}

The integral field spectrograph is one of the three 
operational modes
available on VIMOS (Le F\`evre et al. 2003). 
The IFU consists of 4 quadrants of 1600 fibers each, 
feeding four different 2k$\times$4k CDDs.
Each quadrant is made by four sets ({\it pseudo-slits}) 
of 400 fibers.
Up to date, the VIMOS-IFU is the
integral field spectrograph with the largest field of view (f.o.v.), among
those available on 8/10-m class telescopes: 
the f.o.v. covers (in the lower spatial resolution mode) 
a contiguous sky region 
of $54''\times54''$, 
with 6400 fibers of $0''.66$ diameter. The dead space between adjacent
fibers is less than 10 \% of the fiber dimension. 

The galaxy cluster Abell~2667 has been observed 
in service mode on the nights 29 and 30 June 2003
(as part of program ID: 71.A-3010, PI: Soucail).
Both nights were photometric, with the DIMM seeing varying between
$0''.6$ and $0''.8$ during the observations of the cluster, 
performed at airmass always lower than 1.28.
%
We have used the 
spatial low-resolution mode (fibers with diameter $0''.66$),
and the low-resolution blue grism (LR-B)
in combination with an order sorting filter, 
which covers the wavelength range from 3500 \AA \, to 7000
with spectral resolution $R\sim250$ and dispersion 5.355 \AA /pixel.
However, because of the overlapping of 
the spectra between contiguous pseudo-slits
on the CCD, the first and last $\simeq 50 $
pixels on the raw spectra from most of the pseudo-slits are not usable.
This reduces the useful spectral range approximatively
 from 3900  to 6800 \AA.

The overall exposure time is 10.8 ksec (4 $\times$ 2700 s, 
two exposures were obtained each night).
A small offset of about 2 arcsec among consecutive
exposures has been applied, in order to compensate for the effect 
due to the not uniform efficiency of the fibers and the presence 
of a small set of low quality fibers.

Calibration frames 
have been obtained soon after each one of the 4 exposures and a spectrophotometric
standard star has been observed each night.
The final area is $54''.8 \times 55''.4 $, 
corresponding
to a region of about 200 $\times$ 200 kpc$^{2}$,  
centered $5''$ SW from the brightest cluster galaxy.

\section{Reduction of the VIMOS-IFU data}

The whole reduction process of the VIMOS-IFU data 
has been completed using the VIMOS Interactive Pipeline Graphical Interface
(VIPGI, Scodeggio et al. 2005), 
a dedicate tool to handle and reduce VIMOS data\footnote{VIPGI has been 
developed within the VIRMOS Consortium. See the 
VIPGI web site for more information: 
\texttt{http://cosmos.mi.iasf.cnr.it/marcos/vipgi/vipgi.html}}.
See also Zanichelli et al. (2005) for a description of 
the data reduction methods and the quality assessment.
Every reduction step before the final combination of the 
dithered  exposures in a single data cube
is performed on a single quadrant basis. 
The main steps are the followings: 
check and adjustment of the so-called {\em first guesses}
of the instrumental model (see later on), 
creation of the spectra extraction tables at each pointing, 
CCD preprocessing, wavelength calibration,
cosmic ray hits removal, determination of fiber efficiency,
sky background subtraction and flux calibration.

%
%
The VIMOS-IFU data reduction 
requires a highly accurate description of the 
optical and spectral distortions of the instrument,
especially for the spectra location and the wavelength calibration.
These distortions are modeled using third order polynomials, whose
coefficients (as periodically determined by the staff at the telescope)
are stored in the raw FITS files headers.
However, since these distortions may change in time 
(because of, e.g., the different orientation of the instrument
during observation or the instrument aging)
such a predefined model can only be used as a
first guess when calibrating the scientific frames.
Moreover, we have experienced that most of the times 
these first guesses are not
close enough to the actual instrument distortions
to be safely used, mainly because of the 
large  flexures of the instrument 
between the time of their definition and the time of observation
(see, e.g., D'Odorico et al. 2003).
These can sum up to a few pixels  (see, e.g., D'Odorico et al. 2003),
and are larger for observations at the meridian and close to the zenith, 
like in the present case.
Therefore, the original first guesses
for the spectra location and the inverse dispersion law
need to be checked and corrected for any given pointing,
by using the calibration frames as close in time as possible
to the scientific exposures
and taken at the same rotator absolute position.
At this aim, 
we used a specific graphically guided tool provided by VIPGI
to correct interactively the polynomial coefficients 
describing the optical and wavelength distortions (Scodeggio et al. 2005),
independently for each calibration set associated to the 
scientific exposures.
These corrected values are then used in the following location of spectra 
traces and wavelength calibration.

%
%
Having adjusted the first guesses for the optical distortions,
it is then possible to trace
the spectra of the 4 detectors.
Location of the spectra traces is a very critical step, 
since spectra are highly packed on the CCDs: 
distances between spectra from contiguous fibers is 5 pixels,
each fiber having a FWHM of $\simeq 3.2 $ pixels. 
Therefore, even small errors ($\sim $ 1 pixel) in tracing the exact 
position of the spectra can
result in a degraded quality of the final result.
Because of their high S/N, 
we have used flat field lamps taken immediately after each science exposure 
to trace the spectra.
In this step an extraction table is created, 
which is then used to trace spectra
of the scientific exposures.
The accuracy of the extraction tables has been visually checked 
on the raw science frame themselves, by verifying 
that the fibers with highest signal are indeed correctly traced.

Because of the high density of the spectra on the VIMOS detectors, 
there is a not negligible amount of crosstalk,
i.e., flux contamination among nearby fibers. 
However, we have done no attempt to correct for this effect:
indeed, correcting for the cross-talk 
would need a very good modeling of 
the fibers profile in the cross-dispersion direction.
Zanichelli et al. (2005) have shown that 
the average contamination on each neighboring fiber is 
$\sim 5 \% $, 
i.e. smaller or of the same order of magnitude of the error introduced
by fitting the fibers profile, since the quality of crosstalk correction 
decreases rapidly 
as soon as the error on the width measurements is of the order
of $\simeq 0.2$ pixels.

\begin{figure}
\psfig{file=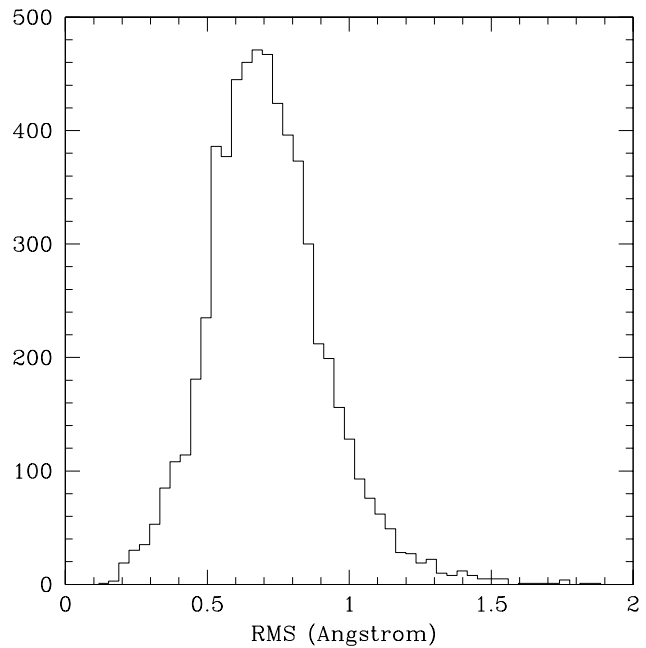,width=0.5\textwidth}
\caption{Distribution of the RMS residuals 
in the wavelength calibration for all the useful fibers in the first pointing.
The median value is 0.716 \AA, i.e. 0.15 pixel.
95\% of the fiber has RMS lower than 1.0 \AA. }
\label{fig:lambda-rms}
\end{figure}

The inverse dispersion law is calculated starting from the 
adjusted first guesses and a fit by a third order polynomial
function. 
All the inverse dispersion solutions have been visually checked, 
by comparing the predicted positions of 
arc lamp lines (Neon and Helium lamps) 
with their effective positions on raw frames.
To demonstrate the accuracy of the wavelength calibration, 
in Fig.~\ref{fig:lambda-rms} the distribution
of the RMS residuals in the
wavelength calibration for the 6400 fibers
in a single pointing is plotted: 
the median values is 0.716 \AA, i.e. about 0.15 pixels,
with negligible differences among the four quadrants.

%
%
The scientific exposures have been 
overscan trimmed and  bias subtracted,
and cosmic ray hits have been cleaned.
The cleaning algorithm (described in Zanichelli et al. 2005)
is based on a sigma-clipping method and 
the fact that along the dispersion direction also spectra with
strong emission lines
show a smooth behavior, while comics rays show very strong gradients.
This method is very efficient in removing hits spanning only a few pixels 
($\sim$ 99\%), and has lower efficiency ($\sim 90$\%) 
just in the case of more
extended ones. A further cleaning has therefore been applied 
in the final combination of the dithered exposures.

Finally, the wavelength calibration is applied 
and the 1D-spectra are extracted.
Extraction of the spectra is based on the Horne's (1986) recipe.
At this stage sky lines are used to 
check and refine the wavelength calibration, in order to compensate the effects
of possible further 
differential flexures between the lamp and scientific exposures. 

Then, the correction of the fiber-to-fiber relative transmission
(analogous to the flat-fielding in the imaging case) is
done by measuring in each 1D spectrum the 
flux contained in an user specified strong sky line, 
either the 5577~\AA \, or the 5892~\AA \, lines. 
We have calculated the relative transmission of each fiber 
independently in 16 different exposures  
with the given grism
performed in the present observing run, and used its median value.

%
Since VIMOS-IFU has not a dedicated set of fibers to determine 
the sky background level,
sky subtraction 
is performed in a statistical way:
in each module,  
fibers are grouped in three sets according to their shape
(as characterized by the FWHM and skewness of the fiber output on the CCD), 
and the sky level is obtained by their mode (Scodeggio et al. 2005).
This approach gives robust 
results when applied to a field in which most of the fibers 
have only sky signal, like in the present case.

Flux calibration is done separately for each quadrant 
of the single exposures, using the observations of a standard star.
A sequence of observations with the star
centered respectively on each quadrant is taken each night.  
From the comparison of the flux-calibrated spectra
with the B$_{450}$ and V$_{606}$ magnitudes, we estimated 
the accuracy of the spectrophotometric calibration to be $\sim 20 \% $.

Finally, the four fully reduced exposures have been combined together,
by using shifts of integer number of fibers. 
The final data cube 
has been corrected for the effect 
due to the differential atmospheric refraction
using the formula from Filippenko (1982)
and converted to the Euro3D FITS data format (Kissler-Patig et al. 2004).

\begin{figure}
\psfig{file=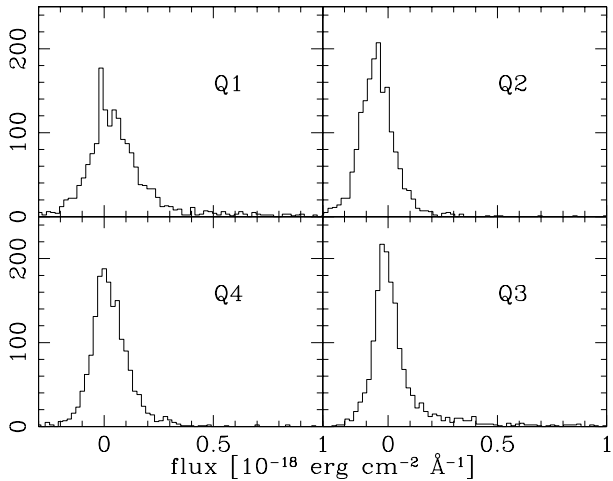,width=0.5\textwidth,angle=270}
\caption{Distribution of the average fibers flux values 
between 4000 and 6200 \AA \, 
in the four quadrants of the fully reduced datacube,
showing the uniformity of the background 
in the different quadrants.
In all quadrants,
distributions are centered around zero, with standard deviation
0.7 - 1.1 $\times 10^{18} \, {\rm erg \, cm}^{-2} \, AA^{-1} \, s^{-1}$.}
\label{fig:flux-distrib}
\end{figure}

The final data cube is made of 6806 spatial elements,
each one representing a spectrum going from $\simeq 3900$ to 
6800 \AA ; it covers a sky area of 0.83 arcmin$^2$, 
centered $5''$ arcsec South-West of the brightest cluster galaxy.
The median spectral resolution is $\simeq 18$ \AA, as estimated from Gaussian
fits to sky lines in the final datacube.
The 3D-cube has been explored and analyzed by using the 
visualization tool E3D (S\'anchez 2004) and specific IDL programs 
written by the authors.
In Fig.~\ref{fig:flux-distrib} we plot the average flux distribution
of all the 6806 spatial elements of the final data-cube: 
globally, the four IFU quadrant show a 
good uniformity in terms of sky background level and noise, 
the flux distribution being very well 
described by a Gaussian peaked around zero, 
with a small tail representing the fibers located on detected objects.
The overall sky background subtraction is relatively accurate, 
as the background level in the final datacube is
around zero in the blank sky regions at all wavelengths.
However, the root-mean-square fluctuations
of the background are a strong
function of the sky position and the wavelength, 
as further discussed in Sect.~\ref{app:1}.
Finally, 
a major limitation above $\sim 6200$ \AA \, is given by the zero order
contaminations, whose position changes from one pseudo-slit to
the other and
from quadrant to quadrant.

\begin{figure}
\psfig{file=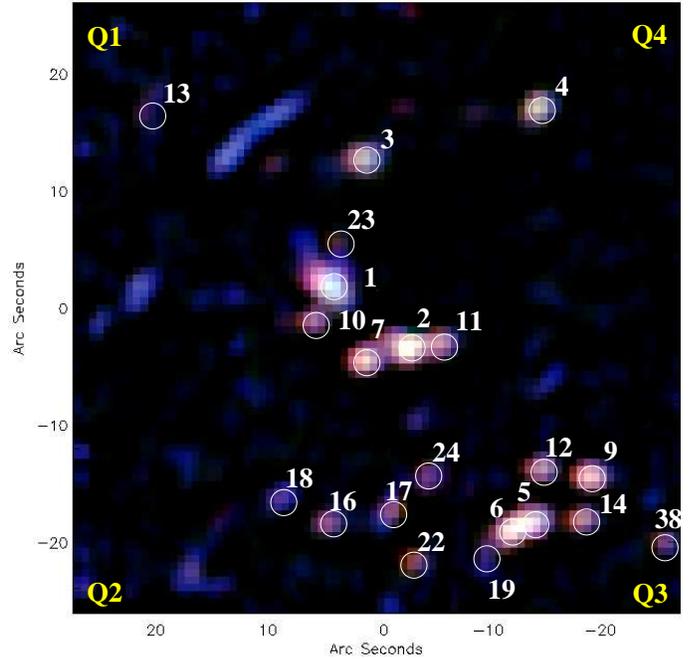,width=0.5\textwidth}
\caption{VIMOS-IFU bi-dimensional color image of the central region 
of the cluster Abell~2667.
The blue, green and red color channels have been built 
by averaging the flux in the spectral ranges:
4500-4700 \AA, 
4900-5400 \AA \, and 5700-6200 \AA , respectively. 
Same orientation as in Fig.~1.
The blue region just NE of the cD galaxy corresponds 
to an extended emission line region at the same redshift of the cD.
The 21 galaxies with secure redshift measurement 
(excluding the lensed sources A and Ra)
are identified in the plot.}
\label{fig:ifu-1}
\end{figure}

A bi-dimensional color projection of the data-cube 
is shown in Fig.~\ref{fig:ifu-1}.
In this image, 
the blue, green and red color channels have been built 
by averaging the flux in the following spectral ranges:
4500-4700 \AA \, (therefore including the [OII] emission 
line at the cluster redshift),
4900-5400 \AA, 
(which covers the 4000 \AA \, break at the cluster redshift), and
5700-6200 \AA.
The giant arc, which is characterized by a strong continuum emission, is 
remarkable at all wavelengths, and 
also the blue emitting region around the central galaxy and the brightest 
cluster galaxies are easily recognized.

\section{The catalog}

\begin{table*}
\caption{Spectrophotometric catalog of the galaxies,  within the VIMOS-IFU f.o.v. 
Columns 4 and 7  list the total magnitudes B$_{\rm 450}$ and  J (AB system),
while colors are measured on $2''$ apertures. Flag represents quality 
of the redshift measurements (see text). 
In the last column, we specify the type of spectral
features identified in the spectra.}
\begin{center}
\label{table:catalog}
\begin{tabular}{r c c c c c c c c l l}
\hline
\hline
ID &  $\alpha {\rm (J2000)}$  & $\delta {\rm (J2000)}$ & B$_{\rm 450}$ & B-V & B-I & J & J-H &  $z$ &   ${\rm flag}$ & {\rm lines}\\
\hline
\noalign{\smallskip}
 1 &23:51:39.4& -26:05:03.3& 19.12& 1.03& 1.90& 16.58&  0.25& 0.2348 &  4 & abs, em \\
 2 &23:51:38.8& -26:05:08.8& 20.34& 1.56& 2.33& 17.32&  0.21& 0.2465 &  4 & abs\\
 3 &23:51:39.1& -26:04:52.8& 20.64& 1.14& 1.66& 18.49&  0.13& 0.2233 &  4 & abs\\
 4 &23:51:38.0& -26:04:48.2& 20.90& 1.24& 1.89& 18.43&  0.14& 0.1570 &  4 & abs\\
 5 &23:51:38.0& -26:05:24.0& 20.90& 1.53& 2.27& 17.94&  0.19& 0.2346 &  4 & abs\\
 6 &23:51:38.2& -26:05:24.8& 20.93& 1.47& 2.20& 18.00&  0.19& 0.2316 &  4 & abs\\
 7 &23:51:39.1& -26:05:10.2& 20.97& 1.60& 2.39& 17.85&  0.23& 0.2316 &  4 & abs\\
 8 &23:51:39.7& -26:04:53.2& 21.06& 1.31& 1.97& 18.73&  0.10& 0.2205 &  1 & abs\\
 9 &23:51:37.6& -26:05:20.0& 21.53& 1.54& 2.32& 18.45&  0.21& 0.2338 &  4 & abs\\
 10&23:51:39.5& -26:05:06.8& 21.73& 1.47& 2.21& 18.79&  0.20& 0.2353 &  4 & abs\\
 11&23:51:38.6& -26:05:08.6& 21.82& 1.55& 2.29& 18.87&  0.20& 0.2309 &  4 & abs\\
 12&23:51:38.0& -26:05:19.4& 21.86& 1.56& 2.31& 18.73&  0.21& 0.2311 &  4 & abs\\
 13&23:51:40.6& -26:04:48.8& 21.89& 1.46& 2.23& 19.00&  0.20& 0.2333 &  2 & abs\\
 14&23:51:37.7& -26:05:23.8& 21.94& 1.51& 2.27& 18.94&  0.19& 0.2325 &  4 & abs\\
 15&23:51:40.3& -26:04:52.1& 21.98& 1.39& 2.10& 19.20&  0.13& --     &  - & \\
 16&23:51:39.4& -26:05:24.4& 22.16& 1.40& 2.12& 19.39&  0.17& 0.2447 &  2 & abs\\
 17&23:51:38.9& -26:05:23.3& 22.22& 1.43& 2.12& 19.44&  0.15& 0.2308 &  4 & abs\\
 18&23:51:39.7& -26:05:22.1& 22.29& 0.82& 1.18& 20.78&  0.04& 0.3980 &  3 & em\\
 19&23:51:38.3& -26:05:27.0& 22.38& 1.31& 1.95& 19.70&  0.17& 0.2290 &  2 & abs\\
 20&23:51:39.6& -26:05:06.7& 22.48& 1.48& 2.20& 19.64&  0.20& 0.2404 &  1 & abs\\
 21&23:51:40.9& -26:05:07.1& 22.55& 0.86& 1.29& 21.11&  0.06& 0.2140 &  1 & abs, em\\
 22&23:51:38.8& -26:05:27.6& 22.60& 1.60& 2.37& 19.47&  0.22& 0.2352 &  4 & abs\\
 23&23:51:39.3& -26:04:59.9& 22.66& 1.43& 2.18& 19.82&  0.20& 0.2360 &  3 & abs, em\\
 24&23:51:38.7& -26:05:20.0& 22.94& 1.50& 2.23& 19.98&  0.18& 0.2351 &  4 & abs\\
 25&23:51:40.2& -26:05:28.6& 23.04& 1.32& 2.02& 20.54&  0.14& 0.2255 &  1 & abs\\
 26&23:51:38.0& -26:04:50.1& 23.40& 1.10& 1.73& 21.04&  0.14& --     &  --& --\\
 27&23:51:37.1& -26:04:53.9& 23.53& 0.30& 0.80& 21.84&  0.17& --     &  --& --\\
 28&23:51:38.4& -26:04:48.6& 23.58& 1.19& 1.88& 21.27&  0.07& 0.2440 &  1 & abs\\
 29&23:51:39.1& -26:04:39.0& 23.59& 0.25& 0.58& 22.12&  0.14& --     & -- & --\\
 30&23:51:38.8& -26:05:15.1& 23.77& 1.51& 2.19& 20.96&  0.12& 0.2336 &  1 & abs\\
 31&23:51:39.3& -26:05:35.2& 23.87& 0.19& 0.16& 22.90& -0.23& --     & -- & --\\
 32&23:51:41.0& -26:04:50.0& 23.89& 0.23& 0.48& 22.66& -0.44& --     & -- & --\\
 33&23:51:39.7& -26:05:01.9& 23.91& 1.39& 2.58& 20.41&  0.05& 0.2380 &  1 & abs\\
 34&23:51:38.0& -26:04:54.6& 24.28& 0.69& 0.99& 23.58& -0.02& --     &  1 & --\\
 35&23:51:39.0& -26:05:24.6& 24.51& 1.42& 2.19& 21.46&  0.10& 0.2270 &  1 & abs\\
 36&23:51:37.3& -26:04:53.7& 24.56& 0.15& 0.40& 23.68& -0.33& --     &  --& --\\
 37&23:51:40.8& -26:05:19.5& 24.81& 1.10& 1.58& 23.22& -0.09& --     &  --& -- \\
 38&23:51:37.1& -26:05:25.3& --   &  -- &  -- & 19.81&  0.20& 0.2301 &  3 & abs\\

\hline
\end{tabular}
\end{center}
\label{table:catalog}
\end{table*}

The photometric catalog of objects within the VIMOS-IFU field of view
has been built using SExtractor (Bertin \& Arnouts 1996) 
on the $\chi^2$-image obtained by combining the 
HST F450W- and F606W-band images
(which overlap the spectral range of the VIMOS-IFU data).
%
%
We have used the Skycat GAIA tool to find the astrometric solution
for the IFU data-cube in order to identify 
the fibers covering the detected objects. 
For each object a 1D spectrum
is obtained by summing the signal on all the 
associated fibers.
%
We have also searched directly the VIMOS-IFU datacube
for objects using a dedicate tool prepared by the authors 
(Covone et al., in preparation).
One bright object (ID = 38) has been detected in the IFU datacube 
and  has no HST counterpart
since this falls outside of the WFPC2 f.o.v. 
Finally, N=39 objects have been detected in the VIMOS-IFU data cube, 
an object being considered as detected if we could find at least a featureless
continuum at the position of the HST object. 

The redshifts have been measured using a cross-correlation technique
by means of the IRAF task {\tt xcsao} (Kurtz et al. 1992),
and attentively visually checked by at least two of us.
The median error on individual measurements, as 
estimated by using this software, is $\sim 0.0003$.
The instrumental uncertainty is about $\sim100$ \kms\ in the
rest frame.
We assigned a confidence class from 1 to 4, 
following Le F\`evre et al. (1995):
class 1 corresponds to a subjective probability of 50 \% that the 
lines identification is correct, 
class 2 to a probability of 75 \% (with more than one spectral feature
identified),
class 3 to a probability higher than 95\% 
and class 4  to 
an unquestionable  identification.

We have obtained redshift measurements for 34 objects, 
including 25 secure ones (i.e., confidence class higher than 1).
The faintest object for which we could measure the redshift has magnitude 
B$_{\rm 450}$ = 24.58 (in a $2''$ aperture).
The final galaxy catalog is shown in Table~\ref{table:catalog},
while information on the gravitationally lensed sources 
is given in 
Table~\ref{table:catalog2}.
Table~\ref{table:catalog} contains photometric and spectroscopic
information for the objects with a redshift measurement, 
and photometric information for the remaining 
objects brighter than
B$_{\rm 450}$ = 24.8. 

\begin{figure}
\psfig{file=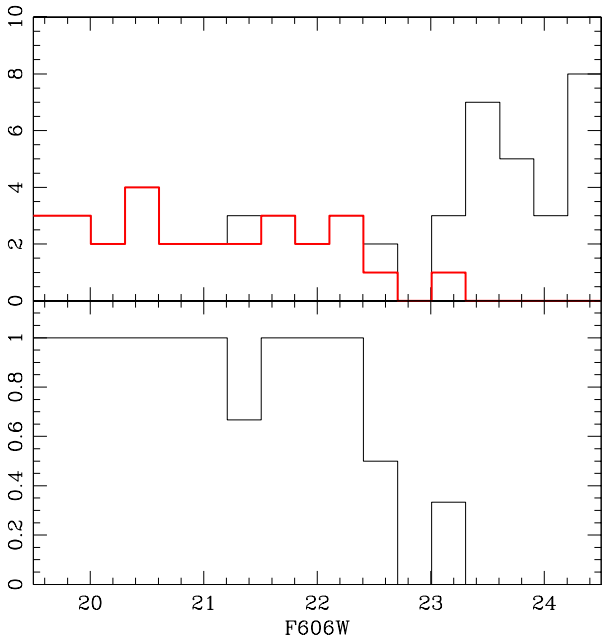,width=0.5\textwidth}
\caption{Completeness of redshift survey as function of V-band magnitude
(measured in a $2''$ aperture).
In the upper panel, we plot the number of objects 
detected on the HST $\chi^2$-image within the VIMOS-IFU f.o.v. (black line),
and the number of those object with measured redshift (red line).
In the lower panel, the efficiency of the spectroscopic survey is given.
The redshift of bright object at $V \simeq 21.5$ (ID=15) 
was not measured since it is located on a set of dead fibers.}
\label{fig:effi}
\end{figure}

\begin{figure}
\psfig{file=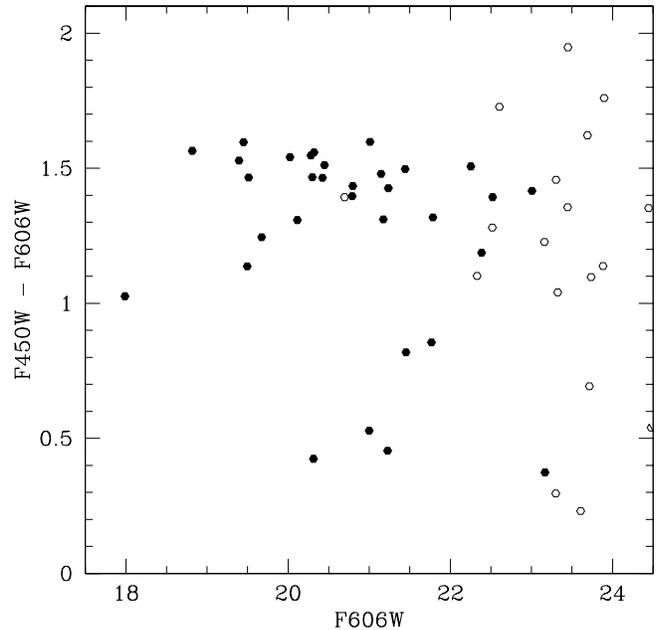,width=0.5\textwidth}
\caption{Color-magnitude relation of  
the galaxies in VIMOS-IFU f.o.v and detected in the HST $\chi^2$-image,
using the two WFPC2 pass-bands covering the LR-B spectral range 
(F450W and F606W). 
Filled symbols correspond to objects with measured redshift,
empty symbols to objects  with no redshift measurement. 
The cD galaxy (ID=1) is the brightest object in the diagram, 
out of the color-magnitude relation 
because of the blue excess associated
with the extended emission line region.}
\label{fig:cm}
\end{figure}

%
%
The overall efficiency of the redshift survey 
within the central region of A2667 is plotted 
in Fig.~\ref{fig:effi}.  
We could measure the spectroscopic redshifts for all galaxies
brighter than B$_{\rm 450} \simeq 23.0$, apart for object \#15,
which is located in a region covered by dead fibers.
The redshift survey is complete down to V$_{\rm 606} \sim 22.5$,
where we could determine redshifts for about half of the objects in the f.o.v.
%
%
In Fig.~\ref{fig:cm} we plot the color-magnitude diagram for all the objects
within the VIMOS-IFU f.o.v. detected on the HST data. 
Redshifts could be measured for 
almost all the galaxies belonging to the cluster red sequence.

\begin{figure}
\psfig{file=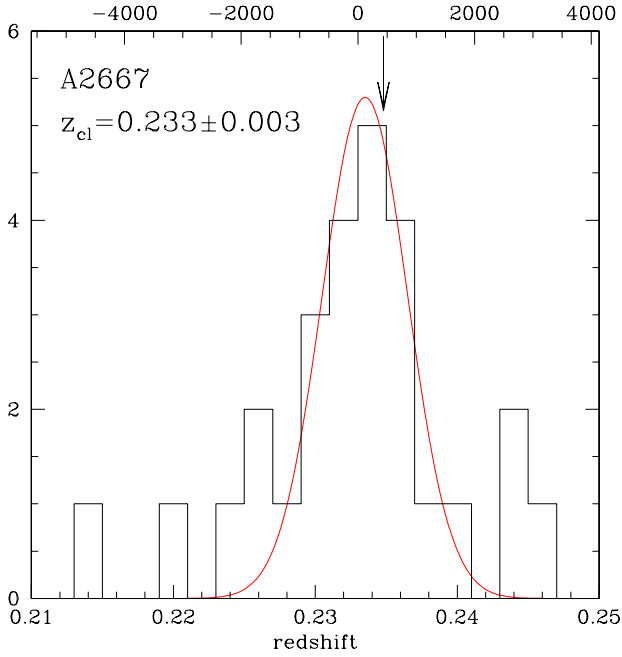,width=0.5\textwidth}
\caption{Redshift distribution 
of galaxies in the vicinity of the cluster.
The bulk of galaxies are bounded to the cluster 
centered at $z=0.233$ with a velocity dispersion of 
960 km s$^{-1}$ (overlayed Gaussian profile).
The arrow marks the position 
of the central galaxy.
Bin size is $\Delta z = 0.002$; on the top axis 
velocities relative to the cluster rest frame are shown (in km s$^{-1}$). }
\label{fig:z}
\end{figure}

%
%
27 galaxies fall in the redshift range $0.21<z<0.25$ (including all 
measurement flags).
We used a classical bi-weight distribution method (Beers et al. 1990)
to determine the cluster membership and velocity dispersion.
Considering 22 galaxies as belonging to the cluster core, 
we derive a cluster redshift of 
$z_{\rm cl} = 0.233 \pm 0.003$ 
(i.e., the median redshift of the assumed cluster members)
and a velocity dispersion of $\sigma_{\rm cl} = 960_{-120}^{+190}$ \kms. 
%
%
The redshift distribution of galaxies in the range $0.21<z<0.25$
is shown in Fig.~\ref{fig:z}.
The cD galaxy redshift is $z=0.2348 \pm 0.0002$, 
which within the uncertainty is
consistent with being at rest in the cluster potential well.

\begin{figure}
\psfig{file=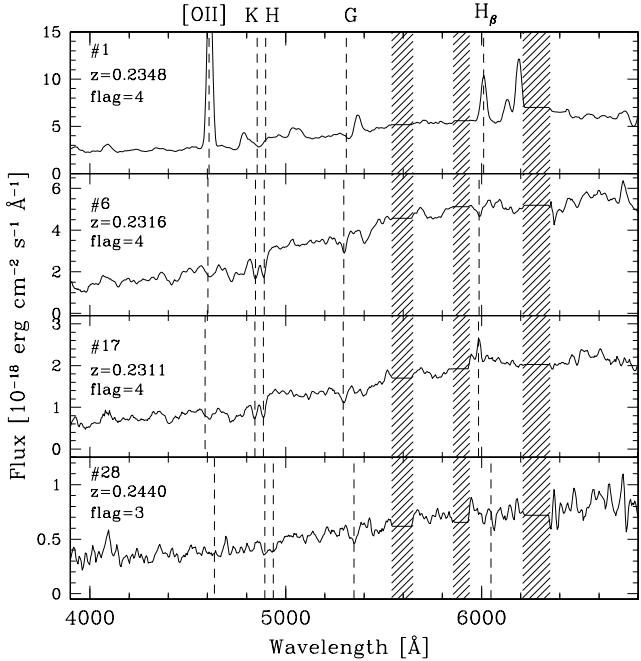,width=0.5\textwidth}
\caption{Four representative spectra of cluster members
from the VIMOS-IFU survey of A2667 core,
in order of decreasing luminosity.
Only a few of the detected lines are indicated.
Spectra have been smoothed at the instrumental resolution.
Shaded region are strongly affected by sky lines residuals.}
\label{fig:spectra-cluster}
\end{figure}

All cluster galaxies show the typical features of evolved early-type
galaxies (a red continuum with strong absorption lines
characteristic of an evolved stellar population), 
as expected in the central region of a relaxed cluster.
Four representative spectra of cluster members are plotted 
in Fig.~\ref{fig:spectra-cluster}.
The largest majority of the
cluster member shows no strong nebular emission line
possibly associated with on-going star formation,
except the cD galaxy which has a rich and 
spatially extended 
structure of strong emission lines.
In particular, the associated [OII] emission line extends well beyond 
the galaxy itself 
(see the blue region visible in Fig.~1 and Fig.~\ref{fig:ifu-1}),
also where the red  continuum of the evolved stellar population
is barely detected.
%
Further investigation of this spatially extended emission and the 
cluster galaxies population will be presented in a forthcoming paper.

\begin{figure}
\psfig{file=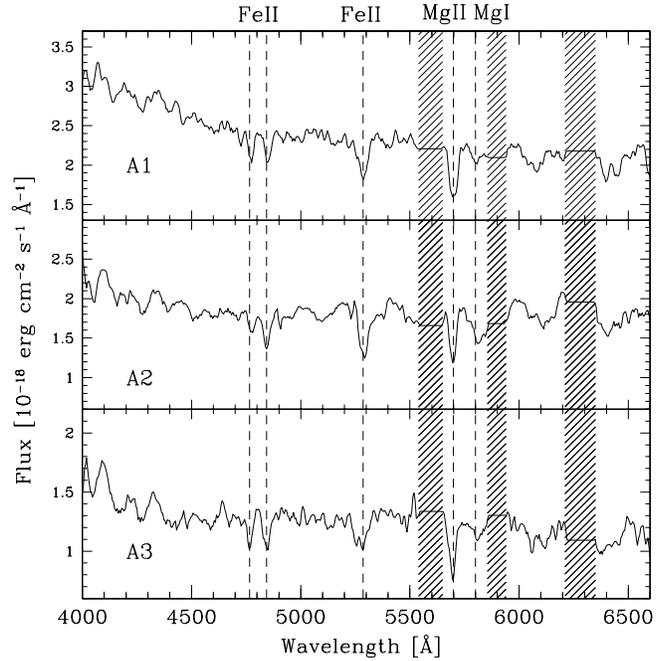,width=0.5\textwidth}
\caption{VIMOS-IFU spectra of the three giant arc components.
Individual spectra are obtained by averaging the flux of the fibers
at the images positions.
Spectra have been box-car smoothed.
Vertical lines show the position of the 
identified UV absorption features.
The redshift of the source is $z_{\rm arc} = 1.0334$.
Shaded regions mark the spectral 
ranges dominated by sky-lines residuals. }
\label{fig:arc}
\end{figure}

In Fig.~\ref{fig:arc} we show the spectra of the three images A1, A2 and A3
forming the giant gravitational arc.
The source is a blue star-forming galaxy
at redshift $1.0334 \pm 0.0003$, whose spectrum is 
characterized by a bright continuum with 
strong UV absorption lines (FeII and MgII),
thus confirming the initial spectroscopic identification of the
[OII] emission line
by Ebeling et al. (in preparation), 
see also Sand et al. (2005).

In Fig.~1 three relatively red, physically unrelated, 
objects are visible:  Ra, Rb, Rc. 
They are not detected in the B$_{\rm 450}$ image
but clearly visible in the redder broad-band images.
They are background galaxies slightly magnified by the cluster gravitational field.

\begin{figure}
\psfig{file=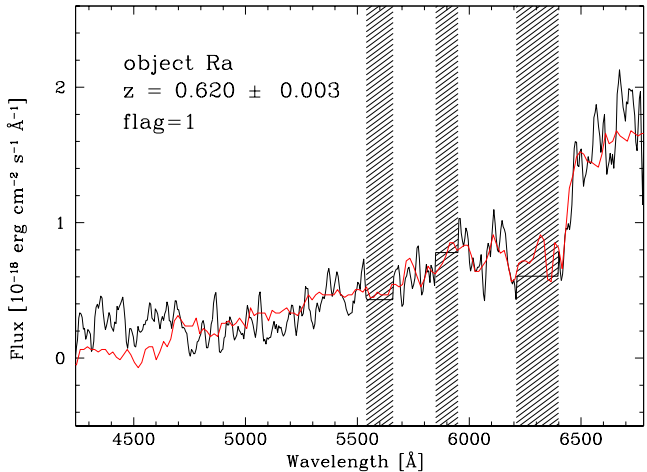,width=0.5\textwidth,angle=270}
\caption{VIMOS-IFU spectrum of the red object Ra
(black line, smoothed at the instrumental resolution), 
together with a redshifted spectrum of a local early-type galaxy 
(red line).
The spectroscopic redshift is based mainly on the 
identification of the 
4000 \AA \, break.}
\label{fig:specRa}
\end{figure}

The spectrum of the object Ra (see Fig. \ref{fig:specRa})
shows a weak red continuum at
wavelengths $\lambda \gtrsim 6000 $ \AA:
the possible identification of the 4000 \AA \, break 
and the overall shape of the continuum
lead to a redshift measurement of $z=0.620$.
The absence of a lensed counterparts allows to put 
a firm upper limit of $z \simeq 3.5$.
 
At the expected position in the VIMOS-IFU data-cube
of the other two red objects (Rb and Rc),
the continuum at $\lambda \gtrsim 6300 $ \AA \,
is too noisy to identify unambiguously any spectral feature.
The absence of lensed counterparts puts an upper limit to Rc
($z <  1.5$), while Rb
is not expected to be multiply imaged at any redshift location.

\begin{table*}
\caption{Properties of the lensed systems and red background galaxies
magnified by the galaxy cluster. 
In the redshift column we list the spectroscopic redshift (if measured)
or the photometric one. 
In the last column, the flag of the spectroscopic redshift
or the predictions $z_{\rm mod}$ 
(or the constraints on the redshift)
from the strong lensing model are given.
Spectroscopic redshifts are obtained by summing all the images in the given
lensed system.}
\begin{center}
\label{table:catalog2}
\begin{tabular}{l c c c  l l}
\hline
\hline
ID  &  $\alpha {\rm (J2000)}$   &  $\delta {\rm (J2000)}$  &  B$_{\rm 450}$  &
redshift  &  notes\\
\hline
\noalign{\smallskip}

{\em multiple images system} \\

A1  &  23:51:39.7 &  -26:04:48.8 & 20.26&  $1.0334 \pm 0.0003$   &  flag=4\\
A2  &  23:51:40.0 &  -26:04:52.0 & 20.62&  $1.0334 \pm 0.0003$   &  flag=4\\
A3  &  23:51:40.6 &  -26:05:04.0 & 21.30&  $1.0334 \pm 0.0003$   &  flag=4\\
\hline 

B1  &  23:51:40.3 &  -26:05:12.0 & 25.13 & $ z_{\rm ph} = 1.20 \pm 0.12$   & 
$z_{\rm mod} = 1.25 \pm 0.05 $\\
B2  &  23:51:40.0 &  -26:05:03.5 & 23.13 & --   & 
$z_{\rm mod} = 1.25 \pm 0.05 $\\
B3  &  23:51:38.3 &  -26:04:44.3 & 25.99 & --   & 
$z_{\rm mod} = 1.25 \pm 0.05 $\\
\hline 

C1  &  23:51:40.6 &  -26:04:56.0 & 24.95 &  $1.578 \pm 0.004$  &  flag=1, \,  
$z_{\rm mod} = 1.6 \pm 0.1 $\\
C2  &  23:51:40.7 &  -26:04:59.1 & 24.86 &  $1.578 \pm 0.004$  &  flag=1, \, 
$z_{\rm mod} = 1.6 \pm 0.1 $\\
\hline 

D1  &  23:51:38.1 &  -26:04:49.8 & 24.29&  $ z_{\rm ph} = 3.12 \pm 0.10 $     & 
$z_{\rm mod} = 3.2 \pm 0.2 $\\
D2  &  23:51:40.3 &  -26:05:19.4 & 25.09&  $ z_{\rm ph} = 2.85 \pm 0.10 $     & 
$z_{\rm mod} = 3.2 \pm 0.2 $\\
&  &  & & \\
{\em  single image systems} &  &  & & \\

Ra  &  23:51:39.4 &  -26:05:03.3 & 23.90&   $0.620 \pm 0.003$  & flag=1   \\
Rb  &  23:51:40.0 &  -26:05:31.5 & 27.09&   $ z_{\rm ph} = 0.95 \pm 0.05 $ & --\\
Rc  &  23:51:37.6 &  -26:05:14.8 & 23.74&   $ z_{\rm ph} = 1.15 \pm 0.05$ &  $ z_{\rm mod} < 1.5 $\\

\hline

\end{tabular}
\end{center}
\end{table*}

\section{Mass models}

\subsection{The lensing mass model and multiple images redshift prediction}

To model the mass distribution of this cluster we used both a 
cluster mass-scale component 
(representing the contribution of the dark matter halo and the intra-cluster
medium) and cluster galaxy mass components 
in a similar way to Kneib et al. (1996), 
see also Smith et al. (2005). 
Cluster galaxies have been selected according to their 
redshift (when available, in the inner cluster region
covered with VIMOS spectroscopy) or 
their J-H color, considering galaxies belonging to 
the cluster red sequence.
We also included the lensing contribution from the foreground galaxy 
\#3 ($z=0.1570 \pm 0.0002$), rescaling its lensing properties 
at the cluster redshift.
In total, 
the mass distribution model is made of 70 mass components, 
including the large scale cluster halo and the 
individual galaxies.
We included in the mass model
all the cluster galaxies brighter than  $\simeq 0.05 \, L_{\rm H}^* $, 
a luminosity limit at which their additional contribution is 
comparable with uncertainties in the overall
cluster model. 
All model components have been parameterized
using a smoothly truncated pseudo-isothermal mass distribution model
(PIEMD, Kassiola \& Kovner 1993), which avoids the unphysical
central singularity and 
infinite spatial extent of the singular isothermal model.

The galaxy mass components have been chosen to have
the same position, ellipticity and orientation 
as of their corresponding H-band image.
Their mass has been scaled with their estimated 
K-band luminosity, 
assuming a Faber-Jackson (1976) relation and a global mass-to-light
ratio (M/L) which is independent of the galaxy luminosity 
(see Appendix in Smith et al. 2005). In short, the K-band luminosity
is computed, assuming a typical E/S0 spectral energy distribution
(redshifted but not
corrected for evolution) for the selected cluster galaxies.
Moreover, we also used a model in which the mass-to-light
ratio has a weak dependence on the
luminosity, $ M/L \propto L^{0.3}$,
as implied by the Fundamental Plane  (Jorgensen et al. 1996,
see also Natarajan \& Kneib 1997). 

Using the {\tt lenstool} ray-tracing code (Kneib 1993), 
we iteratively 
implemented the constraints from the gravitational lenses: 
we started by including the triple imaged 
giant luminous arc A1,2,3. 
The high-resolution HST images show a 
clear mirror symmetry along this giant arc (fold arc)
which gives additional 
constraints on the location of the critical line at the main arc redshift. 
Then we used the predictions of the lensing model
to find additional lensed systems and include their constraints to 
improve the model.
In particular, we used the position and the 
morphology of the following fainter multiple
images: B1, B2, B3; C1, C2, C3 and D1, D2 (see Fig.~1). 
The properties of the gravitationally lensed objects
and other candidate high$-z$
lensed galaxies in this cluster 
are summarized in Table~\ref{table:catalog2}.

Lensing mass models with $\chi^2<1$ are found by fitting the 
ellipticity, orientation, center and mass parameters 
(velocity dispersion, core and  truncation radii) 
of the cluster-scale component 
and the truncation radius and the velocity dispersion
of the ensemble of cluster galaxies 
(using the scaling relations for early-type galaxies).
The best estimates for these parameters are given in Table~\ref{table:model}.
In the following we use mass estimates and redshift prediction from the galaxy
cluster model built assuming a constant M/L 
for the individual galaxies.
However, the predictions of the two models agree 
within the given errors.

In both mass models we find a small offset between the center of the cD
galaxy and the cluster halo component.
This is in agreement with recent findings by Covone et al. (2005),
which have analyzed {\em Chandra} archive data and 
found an offset between the overall X-ray emission
and the central galaxy of less than 1 arcsec.

Finally, the fiducial mass model has been used
to derive constraints on the
redshift of the other detected multiple images: \\
1) The very blue low surface brightness B1-B2 arc 
is predicted at $z=1.25\pm 0.05$, 
which is close to the photometric redshift
derived for B1 ($z_{\rm ph} \simeq 1.2$).
%
In the VIMOS-IFU cube only a noisy spectrum of 
component the two sources 
could be detected, with no useful information to test 
the model prediction. \\
2) The blue pair C1-C2 shows a clear mirror symmetry on the
HST images and is predicted 
to be located at $z=1.6\pm0.1$. 
A possible  identification of CIV in absorption 
would lead to $z = 1.578 \pm 0.002$ (flag = 1), 
very close to value predicted by the strong lensing
model, 
but a higher S/N spectrum is needed in order to confirm the model's 
prediction.
\\
3) Finally, the objects D1 and D2 are likely to be two images of the same 
object at $z=3.2\pm0.2$. The model is predicting two other counter-images,
all of them forming an Einstein cross, but we could not locate these
additional 2 images as they are buried under the cD light. 
Unluckily,
also for this system 
we could not identify any feature in the noisy VIMOS-IFU spectrum
of this object that would have confirmed this redshift prediction, 
although
this time the wavelength coverage was more adequate to search for
the Lyman-$\alpha$ emission line. 
Therefore, this object has probably no
strong Lyman-$\alpha$ emission line.

\begin{table*}
\caption{Parameters of the two fiducial 
gravitational lensing models$^{\mathrm{a}}$,
assuming individual galaxies with  M/L=const and M/L $\propto L^{0.3}$,
respectively.
We list the offset between the cluster dark matter halo
and the cD galaxy, the axis ratio,  the orientation, 
the core radius, the velocity dispersion and
the truncation radius. See text for details.
}
\begin{center}
\label{table:model}
\begin{tabular}{l c c c c c c c }
\hline
\hline
component & $\Delta$R.A.  & $\Delta$Decl. & $a/b$ & $\theta$ & $r_{\rm core}$ & $\sigma_{\rm v}$ 
& $r_{\rm cut}$ \\
          & (arcsec) & (arcsec) & & (deg) &(kpc) &(km s$^{-1}$) & (Mpc) \\
\hline
\noalign{\smallskip}

M/L = const. &&&&&&& \\ 
cluster halo     & -1.6$\pm$0.5 & -0.5$\pm$0.5 & 0.38 & -45 & 74$\pm$8 & 1004 & 1.3        \\
$L_*$ galaxy     & -- & -- & -- & -- & (0.15) &130$\pm$7 & 55$\pm$10   \\
\\
M/L $\propto L^{0.3}$ &&&&&&& \\ 
cluster halo     & --0.8$\pm$0.5 & -1.8$\pm$0.5 & 0.45 & -45 & 76$\pm$8 & 993& 1.3        \\
$L_*$ galaxy     & -- & -- & -- & -- & (0.15) & 155$\pm$8 & 70$\pm$15   \\

\hline
\end{tabular}
\end{center}
\begin{list}{}{}
\item[$^{\mathrm{a}}$]Quantities in parenthesis are not free parameters in the fit.
\end{list}
\end{table*}

\subsection{Comparison of cluster mass estimates}

Our observations and analysis of this cluster allow us to constrain the
mass of the cluster core using both dynamical and lensing mass estimates.
Assuming a singular isothermal sphere model,
the dynamical mass 
can be estimated using the relation (see, e.g., Longair 1998)
\begin{equation}
M(<R) = { \pi \sigma^2 R \over G}.
\end{equation}
Thus for a $30'' $ radius 
($R = 110 \, h_{70}^{-1}$ {\rm kpc}), corresponding
to the area covered by our spectroscopic survey, we derive a mass of
$M = 7.1_{- 1.7}^{+3.1} \, \times 10^{13} \, h_{70}^{-1}$ \, M$_\odot$.

Although the mass within the gravitational 
arc radius is well constrained:
M($<16''$) = $ 2.9 \pm 0.1 \, \times 10^{13}$\,$h^{-1}_{70}$\,M$_\odot$, 
there is some
weak degeneracies in the slope of the mass profile in the cluster core.
Nevertheless, extrapolating to 110\,$h^{-1}_{70}$\,kpc the lensing total mass
is 
$7.2 \pm 0.2 \, \times \, 10^{13} \, h^{-1}_{70} \, M_\odot$, 
a value very close to the dynamical mass estimate. 
%

%
While the present sample of redshift measurements 
does not allow a conclusive analysis of the whole 
cluster dynamical state,
the present result supports the hypothesis 
that the cluster 
inner region is at hydrostatic equilibrium.

Assuming that the cluster follows the $\sigma$--$T_X$ relation 
(e.g., Girardi et  al. 1996), the ASCA observed X-ray temperature of 
$T_X= 5.95_{-0.23}^{+0.42} $ keV (Ota \& Mitsuda 2004) 
corresponds to a velocity dispersion of $1000_{-165}^{+ 190}$ km s$^{-1}$, 
which is in good agreement with both our lensing and dynamical
estimates. 
This also agrees with Allen et al. (1998) findings that galaxy clusters 
with centrally peaked X-ray emission and a short
cooling time generally
show a good agreement between lensing and X-rays mass estimates.

The agreement between the given dynamical and non-dynamical mass estimates
further confirms that the core of A2667 is a relaxed system.
Moreover, we note that the dynamical state of A2667 appears to be
well predicted by its intra-cluster medium (ICM) temperature. 
Indeed, Cypriano et al. (2004) suggested that the ICM temperature 
is a good diagnostic for the dynamical state of a X-ray bright galaxy cluster:
galaxy clusters with X-ray temperature below 
$\sim 8 $ keV are likely to be relaxed systems, hence with a good agreement 
between dynamical and non-dynamical mass estimates 
(see also Cypriano et al. 2005).

However, we note that Allen et al. (2003) X-ray mass model,  
based on the more recent
{\em Chandra} observations, under-predicts the size of the Einstein radius.
Indeed converting the NFW 
parameters of their mass model fit into the
Einstein radius for a $z=1.0334$ source plane we found 
$\theta_{\rm E}$ = 2.7 arcsec, which is far from the observed 16 arcsec.
This discrepancy is not surprising 
because of the small concentration ($c=3$)
of the NFW model.
Indeed, 
keeping the same value of $M_{200}$, a value of $c=$5 -- 6 would match the 
$\theta_{\rm E} = 16'' $ measured.  
Not enough constraints close to the center lead to an
overestimate of $r_s$ and therefore 
to smaller values of the  
concentration $c$ and the  Einstein radius. 
By combining strong and weak lensing Kneib et al. (2003)
and Broadhurst et al. (2005) showed that 
NFW models with large
concentration can match all lensing constraints. 
It is possible that a
combined strong lensing and X-ray analysis would lead to similar results
for A2667, although this is out of the scope of this paper.

Finally, we have determined the rest-frame K-band total light 
derived from the available ISAAC photometry as described in Section
5.1. 
Within the $110 \, h_{70}^{-1} \, {\rm kpc}$ aperture, we found a luminosity of
L$_K=1.2\pm 0.1 \times 10^{12} \, h_{70}^{-2}$~L$_\odot$.
Hence, we derived a M/L$_K$ ratio of $61 \pm 6 \, h_{70}$ M$_\odot$/L$_\odot$. 
This value 
is close to the $z \simeq 0$ values
recently found by 
Rines et al. (2004), which have studied the 
M/L$_{\rm K}$ radial profile 
in a sample of nine nearby, X-ray luminous galaxy clusters.
Within this sample, 
on the scale of $ r \sim 250$ kpc, the measurements scatter in the range 
between 30 and 60 $h_{70} \, {\rm M}_\odot /{\rm L}_\odot$.

\section{Conclusions}

In this work we have presented the first integral field spectroscopic 
survey of a galaxy cluster performed with the VIMOS-IFU,
together with a large set of broad band images of the same field.
We have observed A2667, a massive, X-ray luminous galaxy cluster at 
$z=0.233$,
hosting one of the brightest gravitational arcs in the sky.
We have obtained a 
spatially complete spectroscopic survey 
within the central $\simeq 100 \, h_{70} \, {\rm kpc}$.
The redshift survey is   
complete down to V$_{606} \simeq 22.5$.
We summarize our main findings here:

\begin{enumerate}

\item We have obtained  
spectroscopic redshift measurements for 34 sources, 
including 22 cluster members in the cluster core
and the three images of the gravitational arc ($z=1.0334$).

\item We have used the position and 
measured redshift of 
the giant arc A and 
the position and morphology of the 
multiple images B1-B2-B3, 
the extended triple object C1-C2-C3 and D1-D2
to build a strong 
gravitational lensing model of the matter 
distribution in the central $\sim 200 \times 200 \, {\rm kpc}^2$ around
the cluster center.
The strong lensing models derived using  
two different scaling laws for the 
cluster galaxy population ($M/L=$const and $M/L \propto L^{0.3}$,)
give results which agree within the errors.
The total mass within the Einstein radius ($ r_{\rm E} = 59.6 \, 
h^{-1}_{70} \, {\rm kpc}) $ is 
M($<16''$) = $ 2.9 \pm 0.1 \, \times 10^{13}$\,$h^{-1}_{70}$\,M$_\odot$,
while the extrapolated value at 110\,$h^{-1}_{70}$ \, kpc 
is M = $7.2 \pm 0.2 \, \times \, 10^{13} \, h^{-1}_{70} \, M_\odot$, 
the error bar including 
the weak degeneracies in the slope of the mass profile.
Moreover, the strong lensing models find
a small offset between the dark matter halo and the position
of the cD galaxy ($\Delta =  1.7 \pm 0.5 $ arcsec).

\item We have used the spectroscopic data
for the dynamical analysis of the cluster core, and compared 
our mass estimates with the ones from
recently published X-ray analysis.
Within the radius $R=110 \, h^{-1}_{70}$ \, kpc, the mass 
derived from the dynamical analysis is 
$M = 7.1_{- 1.7}^{+3.1} \, \times 10^{13} \, h_{70}^{-1}$ \, M$_\odot$,
in very good agreement with the 
value from the strong lensing model.
The velocity dispersion is found to be 
$\sigma=960_{-120}^{+190}$ km s$^{-1}$, 
coherent with the value derived from the observed X-ray temperature 
and assuming
that the cluster follows the $\sigma - T_X$ relation.
However, we caution that, although deep, 
our spectroscopic survey is confined to a 
limited linear region of the galaxy cluster,
and a more extended redshift survey is warranted 
in order to define the dynamical status of the galaxy population 
of the cluster as a whole.

\item  The published NFW mass model derived from X-ray measurement
(Allen et al. 2003) finds
a low value for the concentration $c$ 
and can not account for the observed large Einstein radius.
This is likely because of the lack of strong constraints in the 
cluster central region in the X-ray analysis. 
Indeed,  recent studies 
of the lensing clusters ClG~0024+17 and A1689
(Kneib et al. 2003, Broadhurst et al. 2005), 
based on joint weak and strong lensing analyses,
have demonstrated that 
large concentration NFW models are able to 
match all the observables.

\item Within a 110~h$_{70}^{-1}$~kpc radius, 
we find  a rest-frame 
K-band M/L ratio of 61$\pm 5 \, h_{70}$\,M$_\odot$/L$_\odot$,
as expected for an evolved stellar population and  
close to the value found for clusters in the local Universe.

\end{enumerate}

The observed agreement between the dynamical and lensing mass estimates 
supports the idea that the cluster core is in a 
relaxed dynamical state, 
as expected from its regular X-ray morphology 
(Rizza et al. 1998, Allen et al. 2003)
and its ICM temperature (Cypriano et al. 2004).
Therefore, the core of A2667 appears to be dynamically evolved, 
in contrast
with a large fraction ($70 \pm 20 \, \%$)
of similar X-ray luminous clusters at the same redshift 
($z \sim 0.2$, Smith et al. 2005).

This work demonstrates the capabilities of wide f.o.v. 
integral field  spectroscopy 
in obtaining 
spatially resolved spectroscopy
of extended gravitational arcs and 
a complete (spatially and in magnitude) surveys
of the galaxy population in the clusters cores at intermediate redshift.
These measurements are
useful in understanding the cluster mass distribution properties,
using both the strong lensing 
and dynamical analyses.
Wide field IFS is thus a remarkable tool to probe 
compact sky regions such as the cluster cores targeted here.

\begin{acknowledgements}

The authors thank C. Adami, D. Grin, F. La Barbera, S. Brillant,
G.Smith and D. Sand
for useful comments and discussions, A. Biviano for providing 
an updated version of the program {\tt ROSTAT}, and the anonymous referee 
for comments which helped in improving the presentation of the work.
GC thanks his wife, Tina, for the enormous patience and 
invaluable support during all these years.
The data published in this paper have been reduced using VIPGI, developed by 
INAF Milano, in the framework of the VIRMOS Consortium activities.
GC thanks the VIRMOS Consortium team in Milano 
(B. Garilli, P. Franzetti, M. Scodeggio) for the hospitality during his
stays at the IASF and the continuous help on the VIMOS-IFU data reduction.
GC acknowledges support from the EURO-3D Research Training Network,
funded by the European Commission under the contract No. HPRN-CT-2002-00305.
JPK acknowledges support from CNRS and Caltech.

\end{acknowledgements}

\appendix
\section{Sensitivity and flux limits}
\label{app:1}

\begin{figure}
\psfig{file=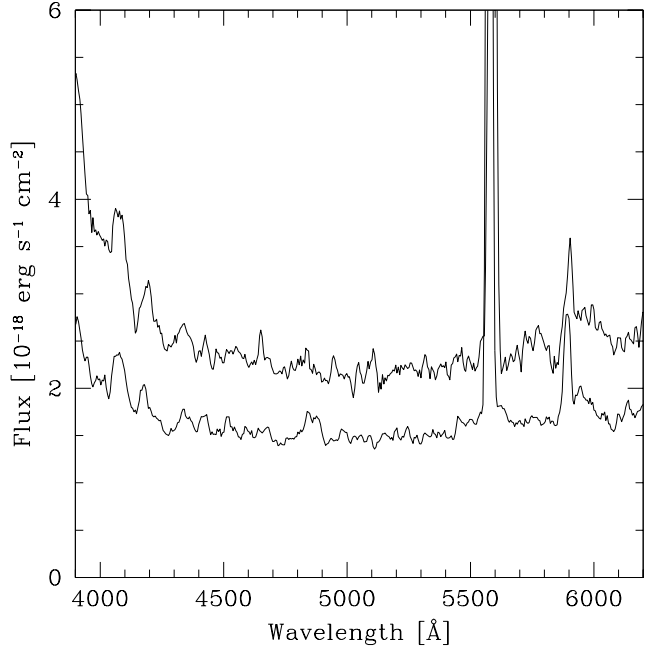,width=0.5\textwidth}
\caption{The flux limits for the detection of emission lines
sources in two different quadrants (Q1, upper curve, and Q2, lower curve)
as a function of the wavelenght. 
The curves for the other two quadrants fall between the two 
plotted here.
See text for details.}
\label{fig:depth:1}
\end{figure}

\begin{figure}
\psfig{file=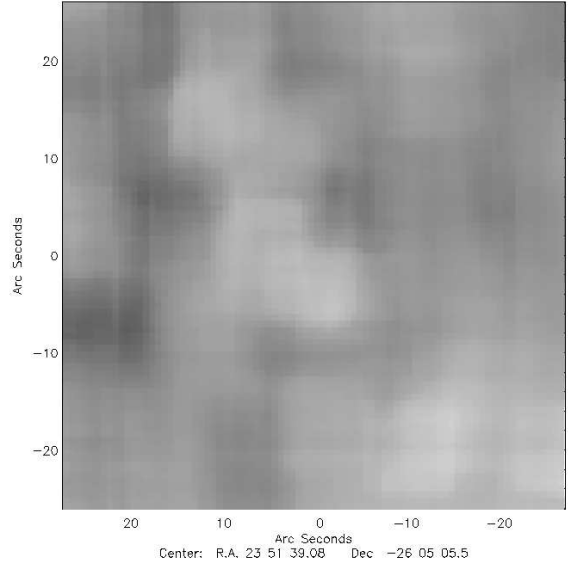,width=0.5\textwidth}
\caption{2D map of the sky background RMS,
considering a window $\delta \lambda = 40 $ \AA 
\, wide at  $\lambda  = 5000 $ \AA .
A logarithmic scale from 
$10^{-19}$ to $10^{-18}$ erg cm$^{-2}$ s$^{-1}$  \AA$^{-1}$
is used, with darker
grey corresponding to lower sky levels.}
\label{fig:skyback}
\end{figure}

While the  VIMOS-IFU observations
presented here
were not meant for a blind search of pure emission-lines objects,
it is anyway interesting to   
evaluate the sensitivity to faint high-$z$ sources
with no detectable continuum, 
i.e. the limiting flux above which a
single emission-line would be detected in our datacube.
See also Santos et al. (2004), for a similar discussion in a long-slit 
search for distant Ly$-\alpha$ sources.

In our datacube, an emission-line could be 
clearly detected if it covers at least 2 spatial elements
and its total flux 
is 3 times the background root-mean-square fluctuations 
(i.e., the sky background noise)
We considered a minimum spectral width of 3 lambda pixels, i.e. about the
spectral resolution.
We plot in Fig.~\ref{fig:depth:1} 
the spatially averaged limiting flux of the whole VIMOS-IFU datacube
as a function of the wavelength.
At the edges of the spectral range, the main contribution comes from the 
worse sensitivity of the whole instrumental device (telescope, instrument,
grism), while in the central regions, where the detection level is more uniform
in wavelength, the limiting emission-line flux 
strongly depends on the atmospheric OH airglow emission.
In order to understand the spatial variations of the sensitivity,
we show in Fig.~\ref{fig:skyback} a bi-dimensional map of the 
defined limiting emission-line flux at $\lambda \simeq 5000 $ \AA. 


We point out that the present observations were not conducted 
with the best strategy: 
an observational sequence of a larger number of
dithered pointings with smaller exposition time
would have been more effective in removing spatial
not-uniformity,
thus improving the final flux limit.

\end{document}